\documentclass[aoas,preprint]{imsart}

\RequirePackage[OT1]{fontenc}
\RequirePackage{amsthm,amsmath,amsfonts,booktabs}
\RequirePackage[authoryear]{natbib}
\RequirePackage[colorlinks,citecolor=blue,urlcolor=blue]{hyperref}
\usepackage{graphicx}
\usepackage{dcolumn,booktabs}
\usepackage{multirow}
\usepackage{bm}
\usepackage{enumitem}
\usepackage{caption}
\usepackage{comment}
\newcolumntype{d}[1]{D{.}{.}{#1}}
\pubyear{0000}
\volume{0}
\issue{0}
\firstpage{0}
\lastpage{0}
\arxiv{arXiv:0000.0000}

\startlocaldefs
\numberwithin{equation}{section}
\theoremstyle{plain} \newtheorem{thm}{{\sc Theorem}}
\theoremstyle{plain} \newtheorem{prop}{{\sc Proposition}}

\theoremstyle{plain} 

\theoremstyle{plain} \newtheorem{lemma}{{\sc Lemma}}

\theoremstyle{plain} 
\theoremstyle{plain} 
\theoremstyle{plain}

\theoremstyle{plain}

\endlocaldefs

\begin{document}

\begin{frontmatter}
\title{\small Hierarchical Infinite Factor Models for Improving the Prediction of Surgical Complications for Geriatric Patients}
\runtitle{Hierarchical Infinite Factor Model}

\begin{aug}
\author{\fnms{Elizabeth} \snm{Lorenzi}\ead[label=e1]{elizabeth.lorenzi@duke.edu}}, 
\author{\fnms{Ricardo} \snm{Henao}\ead[label=e2]{ricardo.henao@duke.edu}}\and 
\author{\fnms{Katherine} \snm{Heller}\ead[label=e3]{katherine.heller@duke.edu}}

\runauthor{E. Lorenzi, R. Henao, and K. Heller}

\affiliation{Duke University}

\address{E. Lorenzi\\
Department of Statistical Science\\
Duke University\\
122 Old Chemistry Building\\
Durham, North Carolina 27708-0251\\
USA\\
\printead{e1}}

\address{R. Henao\\
Department of Biostatistics and Bioinformatics\\
Duke University\\
2424 Erwin Rd, Hock Plaza Suite 1105\\
Durham, North Carolina 27705-3860\\
USA\\
\printead{e2}}

\address{K. Heller\\
Department of Statistical Science\\
Duke University\\
122 Old Chemistry Building\\
Durham, North Carolina 27708-0251\\
USA\\
\printead{e3}}

\end{aug}

\begin{abstract} 
We develop a hierarchical infinite latent factor model (HIFM) to appropriately account for the covariance structure across subpopulations in data. We propose a novel Hierarchical Dirichlet Process shrinkage prior on the loadings matrix that flexibly captures the underlying structure of our data across subpopulations while sharing information to improve inference and prediction. The stick-breaking construction of the prior assumes infinite number of factors and allows for each subpopulation to utilize different subsets of the factor space and select the number of factors needed to best explain the variation. Theoretical results are provided to show support of the prior. We develop the model into a latent factor regression method that excels at prediction and inference of regression coefficients. Simulations are used to validate this strong performance compared to baseline methods. We apply this work to the problem of predicting surgical complications using electronic health record data for  geriatric patients at Duke University Health System (DUHS). We utilize additional surgical encounters at DUHS to enhance learning for the targeted patients. Using HIFM to identify high risk patients improves the sensitivity of predicting death to $91\%$ from $35\%$ based on the currently used heuristic.
\end{abstract} 


\end{frontmatter}

\section{Introduction}
Surgical complications arise in $15\%$ of all surgeries performed and increases up to $50\%$ in high risk surgeries \citet{healey2002complications}. Surgical complications are associated with decreased quality of life to patients and also incur significant costs to the health system. Efforts to address this problem are increasing nationwide with a focus on enhancing preoperative and perioperative care for high-risk and high-cost patients \cite{desebbe2016perioperative}.  Duke University Health System (DUHS) began the Perioperative Optimization of Senior Health (POSH) program, an innovative care redesign that uses expertise from geriatrics, general surgery, and anesthesia to focus on the aspects of care that are most influential for the geriatric surgical population.  POSH developed a heuristic to determine which patients to refer from the surgery clinic visit to their specialized clinic. The heuristic is defined as all patients 85 or older or patients that are 65 or older with cognitive impairment, recent weight loss, multimorbid, or polypharmacy. However, the heuristic identifies about a quarter of the volume of all invasive surgical encounters, which results in more patient visits than POSH can accommodate. %To assist in improving the identification of patients to send to the POSH clinic, we aim to build a risk prediction model that predicts whether a patient will result in complication after surgery. 

Our goal is to identify and characterize high risk geriatric patients who are undergoing an elective, invasive surgical procedure to send to the specialized POSH clinics. We leverage the larger surgical population at DUHS to improve learning, using data derived from electronic health records (EHR). We develop sparse multivariate latent factor models to learn an underlying latent representation of the data that adjusts for the differences between geriatric surgical patients and all other surgical patients, building on the framework introduced by \cite{west2003} and extended by \cite{carvalho2008high},\cite{lucas2006sparse}, and \cite{bhattacharya2011sparse}. Working with high-dimensional EHR data introduces the problems of noisiness, sparsity, and multicollinearity among the covariates. We therefore model the factor loadings matrix as sparse, assuming that only a few variables are related to the factors and therefore the factors represent a parsimonious representation of the data. This modeling approach serves as an exploratory view of underlying phenotypes of the geriatric population compared to the full population that can guide surgeons in deciding which profile of patients would most benefit from interventions. By incorporating response variables, these learned phenotypes are also able to predict post-operative surgical complications with high accuracy.

We focus on modeling the covariance structure of different subpopulations to adjust for the idiosyncratic variations and covariations of each subpopulation. Latent factor models aim to explain the dependence structure of observed data through a sparse decomposition of their covariance matrix. Specifically, factor models decompose the covariance of the observed data of $p$ dimensions, $\Omega$, as $\Lambda \Lambda^T + \Sigma$, where $\Lambda$ is a $p \times k$ loadings matrix  that defines the relationships between each covariate and $k$ latent factors, and $\Sigma$ is a $p \times p$ diagonal matrix of idiosyncratic variances. These models are often used in applications in which the latent factors naturally represent some hidden features such as psychological traits or political ideologies. Others find utility in their use as a dimensionality reduction tool for prediction problems with large $p$ and small $n$ \citet{west2003}. However, our data, derived from noisy EHR data, calls for flexibility beyond the common factor model to better handle the complex structure of the subpopulations we consider, and to induce strong sparsity that can improve predicting outcomes with very low prevalence. A key contribution of this work is the development of the sparse factor model into a transfer learning approach, where we utilize data from a larger source population to improve learning in a target population, in our case the geriatric patients qualified for POSH.

The proposed transfer learning approach places hierarchical priors on the factor loadings matrix. In this setting, we define two groups or subpopulations: the POSH heuristic defined cohort of patients and the remaining invasive procedures occurring at Duke from the entire patient population over the age of 18. There are inherent differences between these two populations. The POSH heuristic identifies a patient cohort with more comorbidities, more medications, and higher complication rates compared to the remaining patients. Modeling these disparate populations requires proper adjustments. Therefore, we introduce a hierarchical infinite prior on the factor loadings matrix, which learns the proper number of factors needed in each group's factor model, while still sharing information across groups to aid in learning for the smaller subpopulation. The hierarchical infinite prior for the factor loadings matrix utilizes the hierarchical Dirichlet process (HDP), a nonparametric model most commonly used within a mixture model, where one may be interested in learning clusters among multiple groups \citet{teh2005}.  The hierarchical infinite prior combines ideas from sparse Bayesian factor models with the hierarchical grouping characteristics of an HDP mixture model, aiming to share information between subpopulations, while capturing the underlying cluster structure, similar to the HDP. We also aim to decompose the sparse covariance structure of our data to model directly the main source of variation between groups, as in a \emph{hierarchical} factor model. We therefore place a hierarchical Dirichlet prior on the loadings matrix of our factor model, $\Lambda$, that flexibly captures the underlying structure of our data across populations.

Section 2 provides necessary background, a detailed overview of our proposed hierarchical infinite factor model (HIFM) utilizing the HDP, and resulting properties of the prior showing that it is well defined and result in a semi-definite covariance matrix. Section 3 presents results in simulations, portraying the properties in model selection, prediction, and interpretability. Section 4 discusses the derived EHR data used to predict surgical complications and review the results. Section 5 concludes and presents future directions for continued effort.

%Tuning the model with an appropriate number of factors is difficult both conceptually and computationally, especially when the goal is to model multiple populations that may require different number of factors to best explain the data. Early work in this area chose the number of factors by maximizing marginal likelihood or different information criteria, such as BIC or AIC. \citet{lopeswest} introduced a reversible-jump MCMC method to learn the number of factors, $k$. \citet{lucas2006sparse,carvalho2008high} choose number of factors by using model selection priors to zero out elements of the loadings matrix. We draw inspiration from \citet{bhattacharya2011sparse}, who proposed allowing the number of factors to approach infinity using a multiplicative gamma shrinkage prior that increasingly shrinks the columns of the loadings matrix to zero. However, their approach has no mechanism for sharing information across groups. This leaves a unique opportunity to develop an infinite factor model that specifically models the subpopulations while sharing across groups.

\section{Hierarchical Infinite Factor Model}
\subsection{Primitives}
The standard Bayesian latent factor model relates observed data, $\bm{x}_i$, to an underlying $k$-vector of random variables, $\bm{f}_i$, using a standard $k$-factor model for each observation, $i \in 1,...,n$ \citet{lopeswest}:
\begin{align}\label{eqn1}
\begin{aligned}
\bm{x}_{i} \sim & N(\Lambda \bm{f}_{i}, \Sigma) \,, \\
\end{aligned}
\end{align}
where $\bm{x}_i$ is a $p$-dimensional vector of covariates, assumed continuous, $\Lambda$ is the $p\times k$ factor loadings matrix where the $j$th row is distributed $\bm{\lambda_j} \sim N(0, \phi^{-1}I_k)$. The k-dimensional factors, $\bm{f}_i$, are independent and identically as $\bm{f}_i \sim N(0, I_k)$, and $\Sigma = \text{diag}(\sigma^2_{1}, ..., \sigma^2_{p})$ is a diagonal matrix that reduces to a set of $p$ independent inverse gamma distributions, with $\sigma_j^2 \sim \text{IG}(a, b)$ for $j= 1,...,p$.  Conditioned on the factors, the observed variables are uncorrelated. Dependence between these observed variables is induced by marginalizing over the distribution of the factors, resulting in the marginal distribution, $\bm{x}_i \sim N_p(0, \Omega)$ where $\Omega = V(\bm{x_i}|\Lambda, \Sigma) = \Lambda \Lambda^T + \Sigma$. Note that there is not an identifiable solution to the above specification and therefore the decomposition of $\Omega$ is not unique. However, for problems involving covariance estimation and prediction, this requirement is not needed, and therefore we do not impose any constraints on the model. This allows us to construct a more flexible parameter-expanded loadings matrix.

We propose a hierarchical stick-breaking prior, motivated by the HDP mixture model formulation. The HDP is hierarchical, nonparametric model in which each subpopulation is modeled with a DP, where the base measure are DP themselves. The DP, $DP(\alpha_0, G_0)$, is a measure on (probability) measures, where $\alpha_0 > 0$ is the concentration parameter, and $G_0$ is the base probability measure \citet{ferguson1973bayesian}. A draw from a DP is formulated as $G = \sum_{h=1}^\infty \pi_h \delta_{\phi_h}$, where $\phi_h$ are independent random variables from $G_0$ and $\delta_{\phi_h}$ are atom locations at $\phi_h$, and $\pi_h$ are the ``stick-breaking'' weights that depend on the parameter $\alpha_0$ \citet{sethuraman1994constructive}. The HDP is a hierarchical model in which the base measure for the children DP are DP themselves, such that:
\begin{align*}
G_0 | \alpha_0, H &\sim DP(\alpha_0, H)\\
G_l | \alpha_l, G_0 &\sim DP(\alpha_l, G_0), \quad \text{for each }\, l.
\end{align*}
This results in each group sharing the components or atom locations, $\bm{\phi}$, while allowing the size of the components to vary per group. It is seen clearly in the following discrete representation, where $\delta_{\phi_h}$ is shared across populations, $l$:
% $G^0 = \sum_{k=1}^{\infty} \pi^0_k \delta_{\phi_k}$, and for each group $G^l = \sum_{k=1}^{\infty} \pi^l_k \delta_{\lambda_k^0}$.
\begin{align*}
    G^0 = & \ \textstyle{\sum}_{h=1}^{\infty} \pi^0_h \delta_{\phi_h} \,, \\
    G^l = & \ \textstyle{\sum}_{h=1}^{\infty} \pi^l_h \delta_{\phi_h} \,, \text{for each }\, l \,.
\end{align*}
\subsection{Proposed Model}
Now consider a $p \times \infty$ loadings matrix, $\Lambda^0$, weighted by the stick-breaking weights of an HDP, such that each population has a unique loadings matrix defined by population specific weights, $\bm{\pi_l}$, where $\Lambda_l = [\sqrt{\pi_{l1}} \lambda^0_1,\sqrt{\pi_{l2}} \lambda^0_2,...]$. The population specific loadings matrix becomes a weighted version of some shared global loadings matrix. The Bayesian factor model prior specification assumes independent rows and columns, so an element in row $j$ and column $h$ from $\Lambda^0$, $\lambda^0_{jh}$, is distributed as a zero-mean normal distribution. Multiplying $\lambda^0_{jh}$  by $\sqrt{\pi_{lh}}$, results in  $\sqrt{\pi_{lh}}\lambda^0_{jh} \sim \text{N}(0, \pi_{lh} \phi^{-1})$. This now mimics the formulation of a scale mixture with the full specification shown in \eqref{prior}, where we represent the prior in the finite case for clarity of the scale mixture specification. The nonparametric process representation is recovered if we let $k \rightarrow \infty$ \citet{teh2005}. We continue with the finite truncation of the model, which is known to be \emph{virtually indistinguishable} from the full process \citet{ishwaran2001gibbs,ishwaran2002approximate}, with $k^*$ as a large \emph{upper bound} for the number of factors. For convenience, we continue to use the notation $k$, where $k$ is sufficiently large. In addition, we set the scale parameter in the loadings matrix, $\phi_{jh}$, constant across populations and distributed gamma in such a way that marginally $\phi_{jh}$ results in a $t$-distribution with $\tau$ degrees of freedom, resulting in a heavy tailed distribution. 
\begin{align}\label{prior}
\begin{aligned}
\bm{\lambda_{lj}}| \bm{\pi_l} \sim & N(0, \bm{\pi_l} \bm{\phi_j}^{-1} I_k) \,, \\
\bm{\pi_l}|\bm{\pi_0} \sim & {\rm Dir}(\alpha_l \bm{\pi_0}) \,, \\
\bm{\pi_0} \sim & {\rm Dir}(\alpha_0/k,...,\alpha_0/k) \,, \\
\phi_{jh} \sim & {\rm Gamma}(\tau/2, \tau/2) \,.
\end{aligned}
\end{align}
The Dirichlet distribution can be decomposed into a set of $k$ independent gamma distributions, such that $w_h \sim \text{Gamma}(\alpha_h, 1)$ for $h = 1,...,k$ and $S : = (w_1 + ... + w_k)$, then $(w_1/S,...,w_k/S) \sim \text{Dir}(\alpha_1,...,\alpha_k)$. We show this for the finite case, but the same is true in the infinite limit, where the Gamma distribution becomes a Gamma process. To induce a closed-form posterior for our proposed prior, we use $k$ unnormalized Gamma draws, $\bm{w_l}$, instead of a draw from a Dirichlet, $\bm{\pi_l}$. The resulting hierarchical prior is specified below in \eqref{wprior}, where $D_{lj} = \text{diag}(w_{l1}/\phi_{j1},...,w_{lk}/\phi_{jk})$.

\begin{align}\label{wprior}
\begin{aligned}
\bm{\lambda_{lj}}| \bm{w_l}, \bm{\phi_j} \sim & N(0, D_{lj}) \,, \\
w_{lh}|\pi_{0h} \sim & {\rm Gamma}(\alpha_l \pi_{0h},1) \,, \quad \forall h \in 1,...,k \,, \\
\bm{\pi_0} \sim & {\rm Dir}(\alpha_0/k,...,\alpha_0/k) \,, \\
\phi_{jh} \sim & {\rm Gamma}(\tau/2, \tau/2) \,,\\
\sigma_j^2 \sim & \text{IG}(a, b)\,.
\end{aligned}
\end{align}
Our prior formulation does not require that $\bm{w_l}$ sums to one, as is the case in a Dirichlet draw. We want the ``rich gets richer" behavior of the HDP that results in many of the stick-breaking weights being approximately zero, signifying the absence of those clusters. By unnormalizing the weights, the same scaling occurs where some weights will be much smaller than others, but now the magnitude is not bounded. This acts as a model shrinkage tool for shrinking factors not needed to describe the distribution of group $l$. We prove a subsequent result in Section \ref{properties}, showing that a loadings matrix with infinite columns results in a finite loadings matrix and covariance structure. The most prominent difference between the HDP and our weighting scheme is that we are not drawing from a discrete measure, instead we use the properties inherent in the stick-breaking process of the sampling proportions to weigh the importance of factors in our model. 

As discussed in \citet{polson2010shrink}, scale mixtures should meet two criteria: first, a local scale parameter should have heavy tails to detect the signal, and second, a global scale parameter should have substantial mass at zero to handle the noise.  Marginalizing over the weights, $\bm{w_l}$, the resulting distribution of $\Lambda$ relates to the normal-gamma shrinkage prior discussed in \citet{caron2008sparse}. To avoid over-shrinking the non-zero loadings, we also define a $k \times p$ matrix of local scale parameters $\phi$ drawn element-wise from a gamma distribution that is constant across populations. This adds an additional source of sharing of information or transfer learning. For example, if an element of the loadings matrix is near zero with small variance, then the signal will also be similar for other subpopulations.

\subsection{Hierarchical Latent Factor Regression}\label{regression}
We utilize the hierarchical infinite factor model to relate the observed covariates to response variables. For each $\bm{x}_i$, we have a corresponding response or a $p_y$-dimensional vector of responses, $\bm{y}_i\in\{0,1\}$. Let $Z=\{Y, X \}$ represent the full data, and the model in \ref{eqn1} simply replaces the $x_i$ with $z_i$.   We concatenate $[f_i, 1]$, and learn an additional column of the loadings matrix. The $k+1$ column of the loadings matrix now serves as an intercept in the model for each covariate. 

The posterior predictive distribution is easily obtained by solving,
\begin{align*}
    & f(y_{n+1}|z_1,...,z_n, x_{n+1}) = \\
    & \hspace{20mm} \int f(y_{n+1}|x_{n+1}, \Theta) \pi(\Theta|z_1,...,z_n) d\Theta \,.
\end{align*}
The joint model implies that $E(y_i|x_i) = x_i^\prime \theta_{l}$ with covariance matrix $\Omega_{l,YX}$, where $\Omega_{l,YX}$ is a partitioned covariance matrix defined by the rows and columns corresponding to $Y$ and $X$. The resulting coefficients, $\theta_{l} = \Omega_{l,XX}^{-1} \Omega_{l,YX}$, are found by correctly partitioning the covariance matrix, $\Omega_{l}$. This then results in the true group-specific regression coefficients of $Y$ on $X$. 

In our application, the data are both binary and continuous, with all outcomes being binary indicators of surgical complications. Therefore, we extend this method to deal with this data structure by using the common probit transformation \citet{albert1993bayesian}. We choose this transformation due to its ease in computation and implementation. With the probit transformation, we convert our binary data to the real line where it now mimics a Gaussian likelihood as the continuous variables do under our model specifications, except in this case we do not learn the idiosyncratic noises, $\Sigma$, and instead set those to 1.

The resulting factor scores represent a transformed feature space of our data that aim to minimize the distributional differences between the populations. Therefore, we proceed with prediction by learning factor scores for the held-out test set of interest. Specifically, we will draw $p(f_i| x_i,\Lambda_{XX}, \Sigma_{XX})$ for each $i$ in the testing set from the defined full conditional for $f_i$, where we subset the learned parameters appropriately to match the testing predictors.

\subsection{Properties of the shrinkage prior}\label{properties}
We let $\Pi_\Lambda \otimes \Pi_\Sigma$ be the prior specification defined in \eqref{wprior}. Because $\Pi_\Lambda$ defines the prior on the infinite dimensional loadings matrix, we must assure that a draw from the prior is well defined and that the elements of the $\Lambda \Lambda^T$ are finite for a semi-definite covariance matrix. As shown in \citet{bhattacharya2011sparse}, we can define a loadings matrix, $\Lambda$, with infinitely many columns while keeping $\Lambda \Lambda^T$'s entries finite. We follow the steps taken in their paper to prove similar properties for our hierarchical infinite factor loadings prior. 

We first define $\Theta_\Lambda$ as the collection of matrices $\Lambda$ with $p$ rows and infinite number of columns, such that the $p \times p$ matrix, $\Lambda \Lambda^T$, results in all finite entries:
\begin{align}\label{finitelam}
\Theta_\Lambda = \{ \Lambda = (\lambda_{jh}), j= & 1,...,p, \\
h= & 1,...,\infty,\max_{1\le j \le p} \sum^{\infty}_{h=1} \lambda^2_{jh} < \infty \} \,. \nonumber
\end{align}
The entries of $\Lambda \Lambda^T$ are finite if and only if the condition in (\ref{finitelam}) is satisfied,  which is possible using the Cauchy-Schwartz inequality and proved in Appendix \ref{App}. All proofs for subsequent properties are shown in Appendix \ref{App}.

Next, let $\Theta_\Sigma$ denote the $p \times p$ diagonal matrices with nonnegative entries and let $\Theta$ denote all $p \times p$ positive semi-definite matrices, and allow $g: \Theta_\Lambda \times \Theta_\Sigma \rightarrow \Theta$ corresponding to $g(\Lambda, \Sigma) = \Lambda \Lambda^T + \Sigma$.  We next define Proposition \ref{prop1} to show that our prior is an element of $\Theta_\Lambda \times \Theta_\Sigma$ almost surely. This reduces to a proof of $\Pi_\Lambda(\Theta_\Lambda) = 1$ under the independence assumption on $\Theta_\Lambda \times \Theta_\Sigma$, where $\Theta_\Sigma$ is well defined as a product of $p$ inverse-gamma distributions. 

%\begin{lemma}\label{lem1}
%For any $(\Lambda, \Sigma) \in \Theta_\Lambda \times \Theta_\Sigma$, the function $g(\Lambda, \Sigma) \in \Theta$.
%\end{lemma

%
\begin{prop}\label{prop1}
If $(\Lambda, \Sigma) \sim \Pi_\Lambda \otimes \Pi_\Sigma$,\, then $\Pi_\Lambda \otimes \Pi_\Sigma(\Theta_\Lambda \times \Theta_\Sigma) = 1$.
\end{prop}
We also show that the resulting posterior distribution of the marginal covariance, $\Omega = \Lambda \Lambda^T + \Sigma$, is weakly consistent by proving Theorem \ref{th2}, defined below:
\begin{thm}\label{th2}
Fix $\Omega_0 \in \Theta$. For any $ \epsilon > 0$, there exists $\epsilon^* >0$ such that:
\begin{align*}
    \{\Omega: d_\infty(\Omega, \Omega_0)<\epsilon^*\} \subset \{\Omega: K(\Omega_0, \Omega) < \epsilon\} \,.
\end{align*}
\end{thm}
Our infinite hierarchical prior meets these properties for each group's estimated covariance by first showing that the prior has large support and therefore places positive probability in $\epsilon$-neighborhoods around any covariance matrix. 

Lastly, we make an argument that the resulting covariance decomposition mimics the results from an HDP mixture model with cluster-specific covariances. For group $l$,  $\Omega_l$ is the population-specific covariance structure of the data, $X$, where $\Omega_l = \Lambda_l \Lambda_l^T + \Sigma_l$.  If we rewrite $\Lambda_l$ as $(\Lambda_0 W_l^{1/2})$ where $W_l$ is diagonal matrix of elements $\bm{w_l}$, we see that the resulting decomposition is  $(\Lambda_0 W_l \Lambda_0^T) + \Sigma$. We then can reformulate this as a sum up to $k$, resulting in a linear combination of rank-1 covariance matrices.
\begin{align}
 \Omega_l &= \Lambda_l \Lambda_l^T + \Sigma_l \\
  &= (\Lambda_0 W_l \Lambda_0^T) + \Sigma_l\\
  &= \sum_{h=1}^k w_h (\bm{\lambda_{0h}} \bm{\lambda_{0h}}^T) + \Sigma_l \,.
\end{align}

\subsection{Inference}
We propose a Markov Chain Monte Carlo (MCMC) scheme with almost all closed-form updates, and provide some suggested updates to allow for faster computation. We truncate the loadings matrix to have $k^* < p$. 

We derive a Gibbs sampler where we draw from the full conditional posteriors. Most posterior updates are derived from conjugate relationships; however, the parameters for the unnormalized HDP are not conjugate. The weight parameters $\bm{w_l}$  are updated with a closed form draw from the generalized inverse-gaussian distribution for each $h$th element of $w_l$:
\[w_{lh}|\lambda_l, \pi_0, \alpha_l \sim {\rm GIG}(p=p_{w_{lh}}, a=a_{w_{lh}}, b= b_{w_{lh}})) \,, \]
where $p_{w_{lh}} = \alpha_l \pi^0_h - p/2$, $a_{w_{lh}}=2$, and $b_{w_{lh}}=(\bm{\lambda_{lh}}'\Phi_h \bm{\lambda_{lh}})$. $\Phi_h = {\rm diag}(\phi_{h1},..,\phi_{hp})$.

To update $\pi_0$, we use a Metropolis-Hastings step within the Gibbs sampler using a gamma proposal with normalization to mimic the Dirichlet distribution. This is done in two steps: First, we propose $\theta_{0h}^* \sim \text{Gamma}(\theta_{0h}^{t-1} \cdot C, C)$, which gives a mean of $\theta_{0h}^{t-1}$ and a variance of $\theta_{0h}^{t-1}/ C$ which allows tuning using the constant, C. We then normalize the $\bm{\theta_0^*}$, such that $\bm{\pi_0^*} = \frac{\bm{\theta_0^*}}{\sum_{h=1}^k \bm{\theta_{0h}^*}}$,
and accept  $\pi_0^*$ based on the acceptance ratio: 
\[ A(\bm{\pi_0^*}|\bm{\pi_0^{t-1}})=\min \left(1,{\frac {P(\bm{\pi_0^*}|w_1,..., w_l)}{P(\bm{\pi_0^{t-1}}|w_1,..., w_l)}}{\frac {g(\bm{\pi_0^{t-1}}|\bm{\pi_0^*})}{g(\bm{\pi_0^*}|\bm{\pi_0^{t-1}})}}\right) \]

All remaining updates from the Gibbs sampler are presented in Appendix \ref{inf}. To speed the computation time of this sampler, we parallelize the updates for the factors $\bm{f_i}$ and the probit transformations of $\bm{x_i}$. Because we assume each row of $\bm{f_i}$ and $\bm{x_i}$ are independent and identically distributed (within each population), we are able to split this update using parallel methods and speed up each iteration by a factor of the number of cores or computing resources present.  

\section{Simulations}
We next evaluate our approach through synthetic data and compare to baseline methods, Lasso and elastic net regressions \citet{tibshirani1996regression}, \cite{zou2005regularization}. Lasso is a commonly used penalized regression model used for variable selection that excels when working with sparse, correlated data, while providing interpretable coefficients that provide insight into the underlying relationships between covariates and outcomes. Elastic net pairs Lasso with ridge regression to share the benefit or both variable selection and regularization, and often results in grouping effects among correlated coefficients. The goal of these analyses is to demonstrate HIFM's capabilities as an interpretable and flexible factor model that excels at prediction. 

We simulate data, $Z_i$, for $i=1,...,1000$ from a $p$-dimensional normal distribution, with zero mean and covariance equal to $\Omega_{l} = \Lambda_l \Lambda_l^T + \Sigma_l$. We simulate with two populations, where 400 observations are within $l=1$, our target. We draw the $j$th row of $\lambda_{lj}$ from a N$(0, D_{lj}^{-1})$, where $D_{lj}^{-1} = \text{diag}(\bm{\phi_{j}}/\bm{w_{l}})$ is a $k \times k$ diagonal matrix. We draw each $\phi_{jh}$ for $j \in 1,..,p$ and $h \in 1,...,k$ from a Gamma($\tau/2, \tau/2$) where $\tau =3$, $w_{lh} \sim \text{Gamma}(\alpha_l \pi_0$,1), and $\pi_0$ from a Dir($\bm{\alpha_0/k}$) with hyperparameters set to $\alpha_l=\alpha_0=15$ to induce approximately uniform clusters. We set the first row of the loadings matrix that corresponds to the outcome, $y$, to 0 and randomly select 2 locations and fill in with a 1 and -1 to induce further sparsity, mimicking a similar design as the simulations presented in \cite{bhattacharya2011sparse}. This adjustment in the simulation aims to make the comparisons across methods fair, with a generative process of the data that is not exactly that of our model. We draw the diagonal of $\Sigma$ from IG(1, 0.33) with prior mean equal to 3.

We compare two different choices of $p$, 50 and 100, with the true number of factors $k=10$. We use the default choice of $5\log(p)$ as the starting number of factors for each simulation run. For each run, we sampled from the Gibbs sampler for 2000 iterations, and remove 1000 iterations for burnin and thinned every fifth iteration. We show two examples: the first with all continuous data as described above, and the second converts the Gaussian simulated data into binary columns using the probit transformation and a random binomial. We convert the first 25 columns, including the outcome, into binary variables for both simulations cases with varying $p$.

We repeat the simulations 50 times and evaluate 1) the prediction performance using an out of sample test set, 2) the precision of the estimated coefficients, and 3) the estimation of the number of factors. We calculate the prediction accuracy for continuous outcomes with mean squared error (MSE), and the binary outcome using Area Under a receiver operator characteristic Curve (AUC), by reporting the median, minimum, and max from the 50 runs. We compare the HIFM model to elastic net and Lasso trained with two different covariate specifications. The first Lasso uses all covariates as main effects and ignores the subpopulation, and the second incorporates a random slope per subpopulation through interactions, which we call a hierarchical Lasso. To tune these models, we use 10-fold cross validation. For lasso, we use the cv.glmnet function from the package glmnet, with their default tuning settings. For elastic net, we cross validate with a grid of 30 parameter settings, where alpha ranges between 0 to 1 in increments of 0.1, and lambda ranges between 0.001 and 1e-5 (using the default tuning grid for lambda from cv.glmnet). 

 Table \ref{all_pred} displays the results from simulations with a continuous outcome, displaying that the hierarchical infinite factor model achieves superior predictive performance compared to elastic net, Lasso and a hierarchical Lasso.  Table \ref{all_pred} also displays the AUC calculated across 50 simulations with binary outcomes, where again HIFM outperforms the alternative models. The baselines provide two gold standards in sparse regression modeling. Elastic net performs slightly better in prediction tasks compared to Lasso and hierarchical lasso, and hierarchical lasso does improve over lasso suggesting the interactions help to better capture the group effects. The improvement by HIFM is significant for both continuous and binary outcomes.

\begin{table}[ht]
\caption{Predictive Performance in simulation study for all simulation cases. Average, minimum, and maximum performance is presented across 50 simulations. Mean squared error (MSE) is calculated for continuous outcome simulations (where smaller is better). Area under receiver operator characteristic curve (AUC) is reported for binary outcomes (where closer to 1 is better). (EN - elastic net, L -lasso, HL-hierarchical lasso.}
\centering
\begin{tabular}{ccccc|cccc}
  \hline
&    \multicolumn{4}{c|}{(50, 10)} &  \multicolumn{4}{c}{(100, 10)} \\
& HIFM & EN & L & HL & HIFM &  EN &L & HL \\
\textbf{MSE} &&&&&&&&\\ 
Mean & 2.82 & 3.18 & 3.41 & 3.31&1.34 & 1.49 & 1.59 & 1.56\\
Min  & 0.13 & 0.24 & 0.23 & 0.17 &  0.12 & 0.19 & 0.21 & 0.18\\
Max &12.42 & 13.28 & 13.73 & 13.78& 16.51 & 16.05 & 15.85 & 15.96\\
%\textbf{MAB}\\ 
%Mean& 1.09 & 1.19 & 1.24 & 1.20& 1.71 && 2.25 & 2.15  \\
%Min & 0.29 & 0.40 & 0.39 & 0.34& 0.58 && 1.07 & 0.92\\
%Max& 2.83 & 2.90 & 2.89 & 2.89 &5.99 && 6.29 & 6.56 \\
\textbf{AUC} &&&&&&&&\\
Mean &  0.81 & 0.74 & 0.71 & 0.72 & 0.81 & 0.76 & 0.73 & 0.72\\
Min  & 0.51 & 0.44 & 0.50 & 0.42 & 0.52 & 0.50 & 0.45 & 0.50\\
Max & 0.94 & 0.93 & 0.92 & 0.94&  0.95 & 0.91 & 0.91 & 0.91\\ 
% \hline 
   \hline
\end{tabular}
\label{all_pred}
\end{table}

Table \ref{coef_perf} displays resulting accuracy of the learned coefficients for each population across method. Coefficients from HIFM are derived from transforming the partitioned covariance matrix of the learned model. We compare the results of HIFM learned regression coefficients to those learned by Lasso with and without interactions and elastic net. We display the results for the simulation of $p=100$ and $k=10$ for both continuous and binary outcomes and averaged over 50 iterations. Similar patterns occurred in the smaller covariate simulation cases, where $p=50$; therefore, we do not report these additional results. The hierarchical Lasso improves the model fit compared to regular lasso, providing evidence that modeling these data hierarchically aids in coefficient estimation. Compared to Lasso and elastic net, HIFM captures the true coefficients with much greater accuracy, for both populations. Interestingly, elastic net performs much worse in the estimation of regression coefficients compared to HIFM and Lasso. The simulation induces very strong sparsity, where the resulting coefficients are very close to zero, especially in the higher dimensional scenario. While lasso may be overshrinking the signal in the data which is why we see worse performance in prediction compared to elastic net and HIFM, the strong shrinkage results in better accuracy across all coefficients compared to elastic net. From these simulations, HIFM shows that it is better at capturing both the coefficient estimates of the data and results in much improved prediction accuracy.

\begin{table}[ht]
\caption{Performance in estimating regression coefficients in simulation study. We report results with $p=100$, $k=10$ for 50 simulations for both continuous and binary examples, showing mean squared error ($\times 10^3$) of estimated coefficients compared to true simulated coefficients.}
\centering
\begin{tabular}{rrrrr|rrrrrr}
  \hline
 & \multicolumn{4}{c|}{Continuous Outcomes} &   \multicolumn{4}{c}{Binary Outcomes} \\
 & HIFM &  EN & L & HL  &  HIFM & EN&   L & HL \\ 
  \hline
Pop 1: &&&&&&&&\\
Median &  0.04 & 1.64 & 1.45 & 0.60 & 0.22 & 12.64 & 2.78 & 1.41\\ 
Min & 0.01 & 0.15 & 0.11 & 0.10 &0.05 & 0.15 & 0.10 & 0.09 \\ 
Max & 9.89 & 16.75 & 18.71 & 16.77& 16.61 & 66.64 & 27.15 & 22.07\\ 
\hline
Pop 2: &&&&&&&&\\
Median & 0.89 & 0.93 & 0.95 & 0.97& 1.11 & 11.90 & 2.55 & 1.93\\
Min & 0.11 & 0.15 & 0.10 & 0.06& 0.10 & 0.11 & 0.06 & 0.06 \\ 
Max & 17.16 & 12.51 & 12.48 & 15.71 & 128.67 & 65.36 & 16.29 & 16.33\\ 
   \hline
\end{tabular}
\label{coef_perf}
\end{table}

\begin{figure}[h!]
    \caption{Visualization of loadings matrix for both simulated populations under HIFM learned with 20 factors. The image plot displays the posterior of the loadings matrix and the scatterplot displays the posterior mean of the weights, $\bm{w_l}$, where the red line indicates the chosen threshold used to determine number of factors in Table \ref{num_factors}.}
    \centering
    \includegraphics[width=\textwidth]{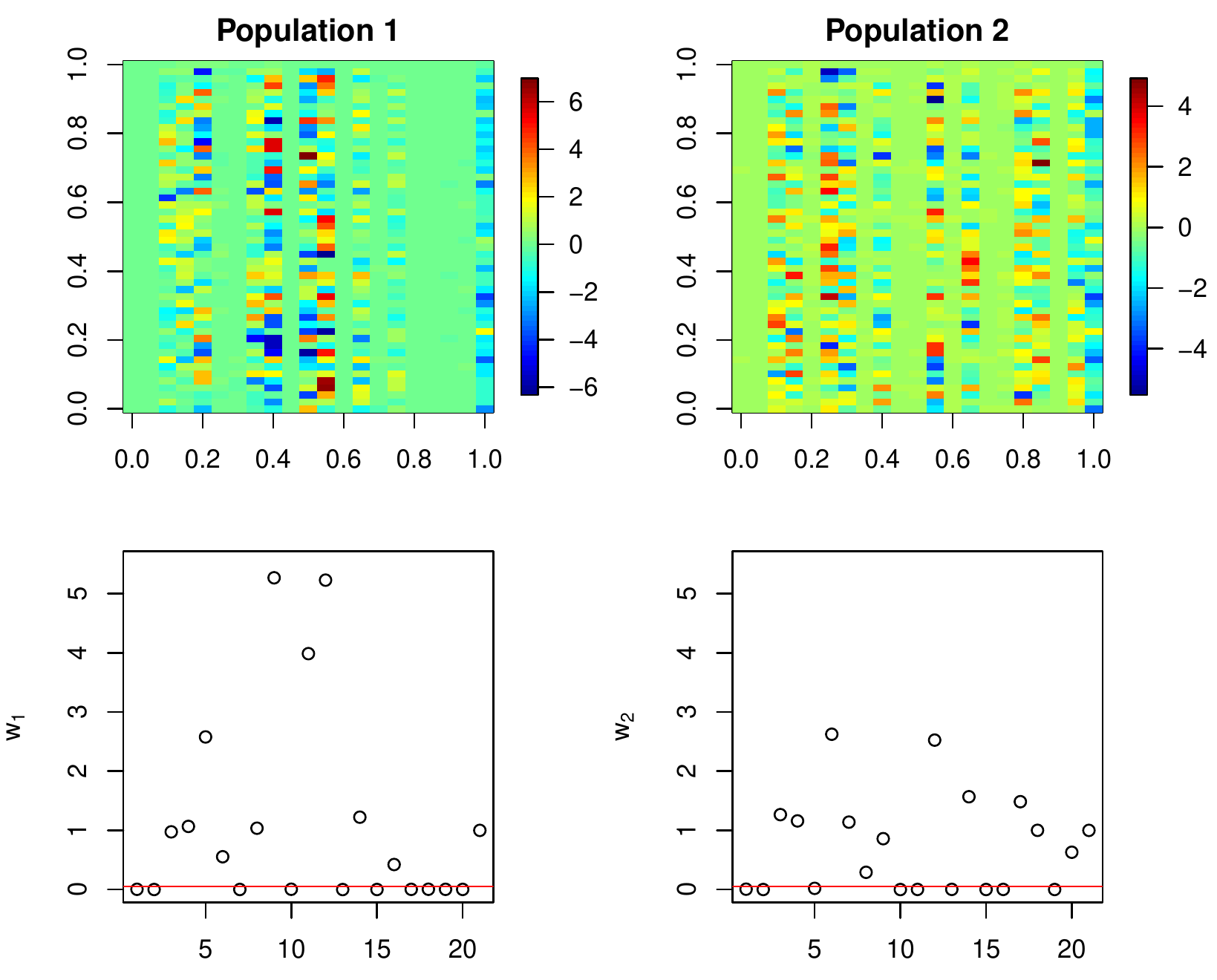}
    \label{image_loadings}
\end{figure}

Lastly, we compare the number of factors used by HIFM for each population and compare those to the true number under simulation. Though we set K=10 in simulation, we incorporate the weights in the loadings matrix that potentially shrink some of the factors across simulations. We set 0.05 as a threshold for considering whether that factor is included or not in the model when evaluating the number of factors chosen. This choice is arbitrary, but the results below were not sensitive to the chosen threshold within a reasonable range. For the HIFM, we set $K=5\log(p)$, where in this scenario $p=50$ so $k$ was set to 20 for the HIFM. We choose to look at the first example with 50 covariates for brevity. From Table \ref{num_factors}, we see that on average HIFM selected 9 factors for each population when all variables were continuous. For binary outcome (and half of all covariates being binary), HIFM selected 9 and 12 factors for first and second population, respectively. The true number of factors simulated averaged at 9 factors for both populations and both types of outcomes, showing that our weighting mechanism was able to recover close to the truth.  Figure \ref{image_loadings} displays the resulting loadings matrix and the posterior mean of the weights post-burn in and thinning for both populations. The model selection properties using the weights are highlighted with the visualization, showing the shrinkage through the weights being used as a model selection tool for the number of factors to include in the model.

\begin{table}[ht]
\caption{Average number of factors selected by HIFM compared to truth. Results displayed for simulations with $p=50$, and $k=10$, with HIFM $k$ set to 20.}
\label{num_factors}
\begin{tabular}{rrr}
  \hline
 & Pop. 1 & Pop. 2 \\ 
  \hline
Continuous Outcome:\\
Normal HIFM & 8.80 & 9.12 \\ 
True & 8.76 & 8.72 \\ 
  \hline \hline\\
Binary Outcome:\\
Normal & 8.82 & 11.94 \\ 
True & 8.80 & 8.90 \\ 
   \hline
\end{tabular}
\end{table}

\section{Surgical Complications}
%\subsection{Data}
\subsection{Goals, Context, and Data}
\begin{figure}[ht!]
\caption{$t$-SNE representation of EHR data from Duke University that meets the POSH heuristic (red) and full patient populations (black), using samples of size 10,000 for each group. Displays low-dimensional projection of full data.}
\centering
\includegraphics[width=.5\linewidth]{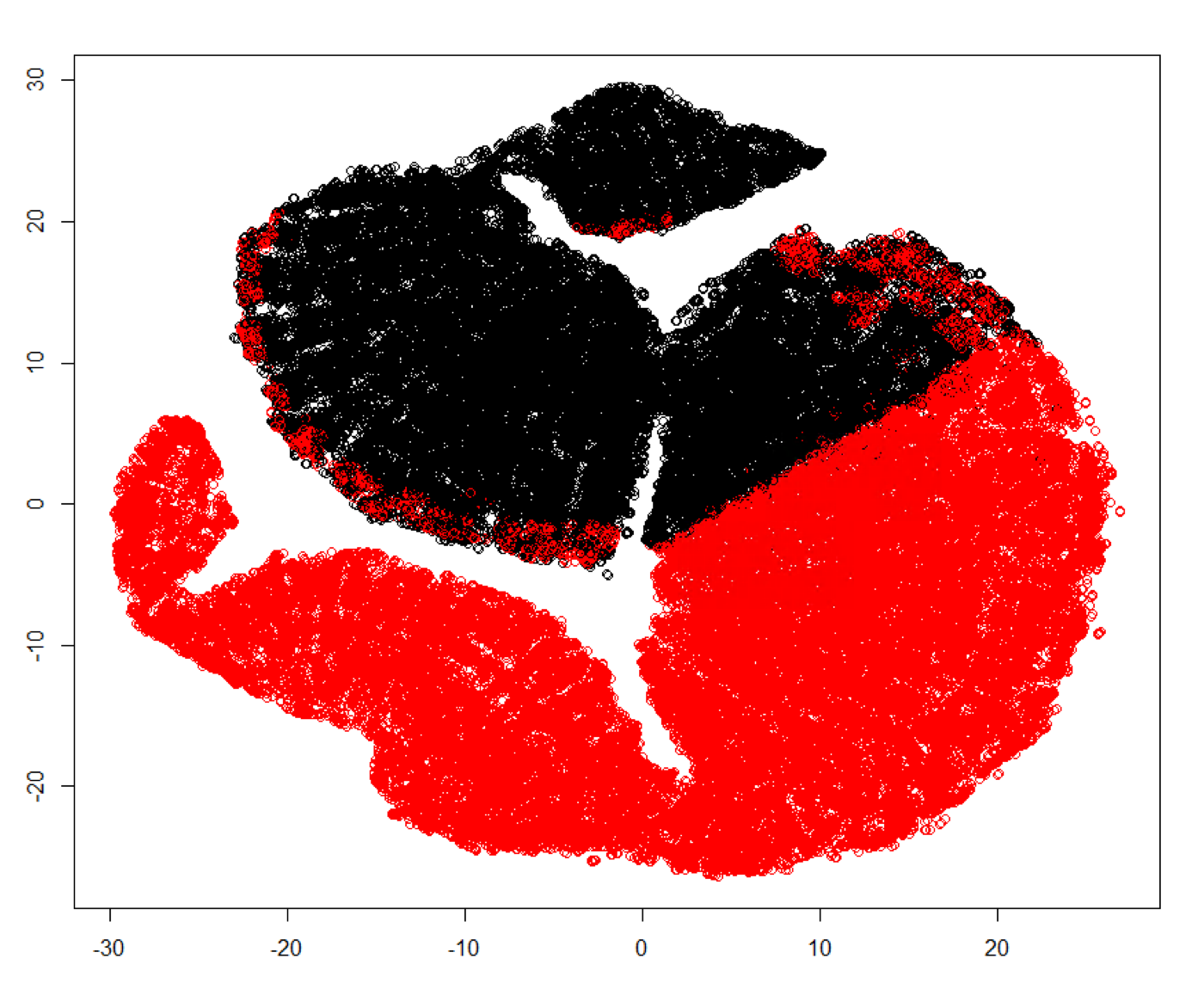}
\label{rtsne_pop}
\end{figure}

Nearly a third of all surgeries performed in the United States occur for people over the age of 65. Furthermore, these older adults experience a higher rate of postoperative morbidity and mortality \cite{etzioni2003elderly} \cite{hanover2001operative}. Complications for older adults may also lead to slower recovery, longer postoperative hospital stays, more complex care needs at discharge, loss of independence, and high readmission rates \cite{speziale2011short} \cite{raval2012outcomes}. The established predictors of poor outcomes such as age, presence of comorbidities, and the type of surgical procedure performed are important predictors for all patient populations, including the geriatric population. However, other factors such as functional status, cognition, nutrition, mobility, and recent falls are less routinely collected factors that are highly correlated with surgical risk among older adults \cite{jones2013relationship}.  This suggests that there are significant differences between the geriatric population compared to the overall surgical population. In Figure \ref{rtsne_pop}, we present the $t$-distributed Stochastic Neighbor Embedding (t-SNE) representation of the Pythia database of all invasive procedures at Duke University Health System (DUHS) between January 2014-January 2017, with samples of 10,000 from the geriatric population that meets the POSH heuristic requirements and 10,000 from the full data. The figure shows patient sub-structure in the data, with a clear difference in the two populations. While there is some overlap between the two populations, it is clear geriatric patients have a different covariate space compared to the overall population. In addition, the figure shows cluster structure which suggests that there also exists natural phenotypes of patients inherent in each group.

The data displayed in Figure \ref{rtsne_pop} is derived from the data repository, Pythia, of electronic health records (EHR) from all invasive surgical encounters from DUHS. Invasive procedures are defined using the encounter's current procedural terminology (CPT) code and included all CPT codes that are identified by the Surgery Flag Software \cite{ahrq}, and eliminated all patients under 18 years of age. Using data derived from the EHR provides the logistical benefit of easier implementation of the resulting tool in a clinical setting since the variables are conveniently found in a patient chart. However, EHR data are a by-product of day-to-day hospital activities, and the resulting data are known to be noisy and sparse. We therefore preprocessed the data to provide a cleaner and more manageable set of covariates to model.  

We include covariates describing the surgical procedure, current medications of the patient, relevant comorbidities, and other demographic information.  The procedure information was captured by CPT codes and grouped into 128 procedure groupings categorized by the Clinical Classification Software (CCS). Procedural groupings with fewer than 200 total patients were removed and grouped into one larger miscellaneous category. This helped to assure that procedural effects were averaged across many patients and represented an overall effect size for geriatric patients and all surgical patients. We defined patient comorbidities by surveying all International Statistical Classification of Diseases (ICD) codes within one year preceding the date of the procedure and classified these diagnoses codes into 29 binary comorbidity groupings (S1) as defined by Elixhauser Comorbidity Index \cite{eshr98}.  We grouped the active outpatient medications recorded during medication reconciliation at preoperative visits into 15 therapeutic binary indicator features and created a separate feature that counted the total number of active medications. We define the outcomes, surgical complications, by diagnosis codes occurring within 30 days following the date of the invasive procedure. The outcomes were derived from 271 diagnosis codes and grouped into 12 categories that aligned with prior studies evaluating post-surgical complications \cite{mcdonald2018}  We use five of these outcomes to focus on the intervention goals of the POSH clinic.  For example, neurological complications encompasses dementia, a common complication for patients over 65 and one that the POSH clinic specifically targets for their patients.  The five outcomes modeled and reported below are cardiac complications, neurological complications, vascular complications, pulmonary complications, and 90 day mortality. Mortality was identified as death occurring within 90 days of the index procedure date. Mortality is captured in the EHR during encounters for in-hospital death and uploaded from the Social Security Death Index for out-of-hospital deaths. Encounters missing EHR data were deemed not missing at random and were therefore excluded from the model development cohort. The resulting covariates are a mix of both continuous (BMI, age, etc.) and binary (indicator of comorbidities, etc), and therefore we utilize the probit transformation that was described above for all binary variables. In addition, we center and scale the continuous variables, and also include an intercept in the model to learn the adjusted mean of the transformed binary variables.

We selected a cohort of 58,656 patients from Pythia that had undergone 77,150 invasive procedural encounters between January 2014 and January 2017 with all complete data.  Of those encounters, 22,055 are flagged as encounters that meet the POSH heuristic determined in clinical practice by surgeons and geriatricians: patient over the age of 85 OR a patient over the age of 65 with greater than 5 different medications, having 2 or more comorbidities, or whether the patient had a recent weight loss or signs of dementia. We form a binary variable to indicate whether a patient meets the POSH heuristic or not, and use that grouping variable to determine the hierarchical structure in the factor model. 

%In addition to comparing our method to the baseline, LASSO, we also compare our model under different training scenarios to show that our method results in positive transfer. 
\subsection{Results}
Our interests are twofold: learn important subset of features and provide accurate predictions of risks of complication for both POSH and all surgical patients. Our goal is to show that pairing the POSH heuristic with a data-driven predictive modeling approach improves the triaging of patients into the high-risk clinic. Additionally, by understanding the covariates that most impact this high-risk geriatric population, we provide insights into the characteristics of the patient that make her/him high risk, and therefore suggests other characteristics to be added to the current heuristic or develop possible interventions to target these characteristics. 

We trained the model on 60,000 encounters from the Pythia database, and held out the 17,149 remaining encounters for validation, of which 4,876 encounter met the POSH heuristic. We ran our Gibbs sampler for 3000 iterations, with burnin of 1500, and thinned every 6 observations. The hyperparameters for the HDP were set to $\alpha_0 = 10$, and $\alpha_1 = \alpha_2 =15$ with the tuning parameter for the Metropolis-Hastings step, C=50.  We set the upper bound for $k$ equal to 25. These settings result in a sparse parameter setting suggesting that many of the factors are shrunk to zero.  

\begin{figure}[t!]
    \centering
    \includegraphics[width=.9\linewidth]{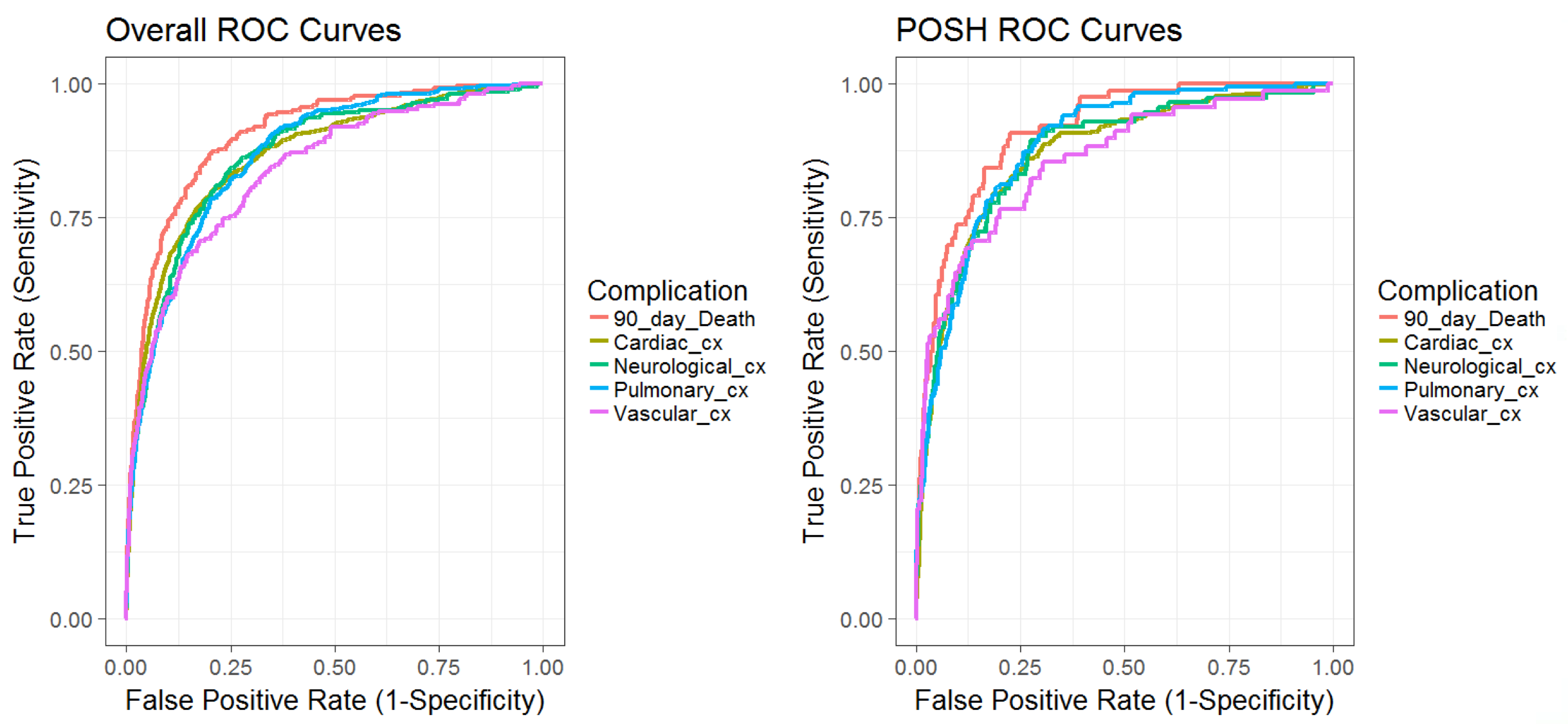}
    \caption{Receiver Operating Curve (ROC) of the five outcomes under the HIFM for encounters across the whole held-out test set and for the test set of geriatric patients. Posterior means with 95$\%$ credible intervals are displayed. } 
    \label{roc}
\end{figure}

To evaluate the predictive performance, we estimated the posterior predictive distribution and evaluated our predicted probabilities compared to the true outcomes. We use the posterior mean of the predictions and calculated the Receiver Operator Characteristic (ROC) curves for both the entire test set and then the POSH encounters within the test-set. Figure \ref{roc} displays the resulting ROC curves. All complications achieved strong performance with AUC between 0.84 - 0.91. Table \ref{pythia_aucs} displays the resulting area under the ROC curves (AUC) and the area under precision-recall curves (AUPRC) comparing the overall test set and the POSH-only test set. We see that the performance is as good and in some cases better in the targeted POSH encounters compared to the full test set. This suggests that our method is able to borrow strength from the larger group to improve the prediction for the smaller targeted group. 

\begin{table}[t!]
\caption{Classification results on five surgical outcomes, comparing full results and POSH specific results for the 5 outcomes. }
\centering
\begin{tabular}{c|cc|cc}
\hline
& \multicolumn{2}{c|}{HIFM - Full}
& \multicolumn{2}{c}{HIFM- POSH}\\
& AUC & AUPRC & AUC & AUPRC \\
\hline
Mortality & 0.905 & 0.192 & 0.901 & 0.187  \\
Cardiac & 0.866 & 0.399 & 0.911 & 0.209\\
Vascular & 0.840 & 0.151 & 0.867 & 0.402 \\
Neurological & 0.864 & 0.172 & 0.868 & 0.408 \\
Pulmonary & 0.867 & 0.246 & 0.872  & 0.148 \\
\hline
\end{tabular}
\label{pythia_aucs}
\end{table}

In addition, we compare the sensitivities and specificities of the resulting model to those of the baseline POSH heuristic. Note that we remove the 500 patients that did go to the POSH clinic from the data so that we do not bias the results with possible treatment effects of the POSH clinics on the patients' outcomes. For the outcome death, the sensitivity and specificity for HIFM are 0.908 and 0.775, respectively. Alternatively, the POSH heuristic achieves a 0.345 sensitivity and 0.716 specificity. The POSH heuristic aims to target high risk patients, not necessarily defined to be high risk of death, though this outcome serves as the best proxy of overall risk. Currently, the POSH heuristic only identified $35\%$ of patients that died, while using the HIFM model in conjunction with the heuristic improves sensitivity to $91\%$, providing evidence that our model is able to effectively identify those patients that are high risk and should go to POSH.

%\end{minipage}
%\begin{minipage}[t]{.4\textwidth}

\begin{figure}
\centering
    \includegraphics[width=1.1\textwidth]{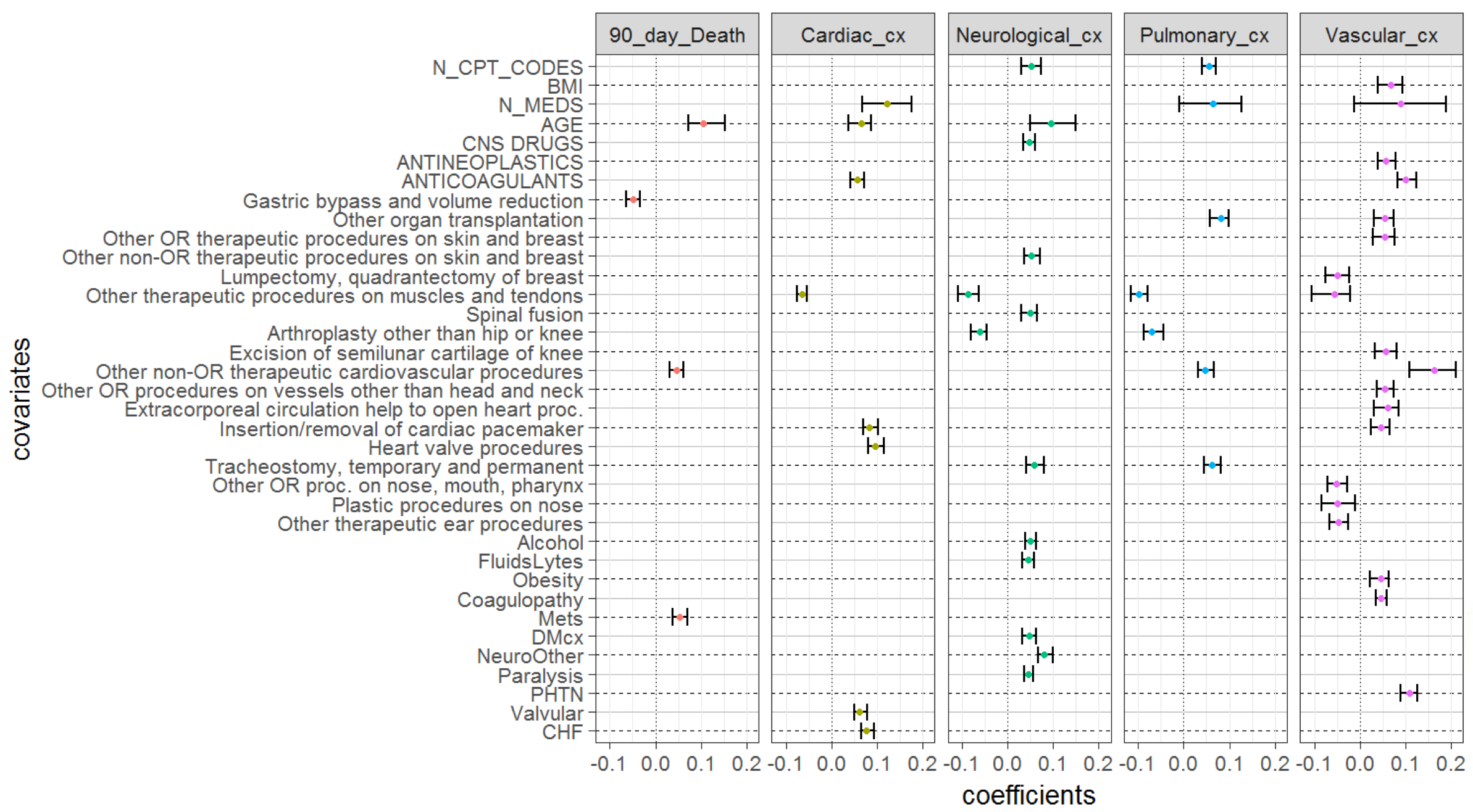}
\caption{Largest estimated coefficients ($|\beta| >=0.05$) for POSH group from HIFM. Posterior means with $95\%$ credible intervals are plotted for each. }
\label{coefs}
\end{figure}

We next calculate the resulting coefficients derived for the POSH specific population through the partitioned covariance matrix discussed in Section \ref{regression}, and find the posterior mean after burn-in and thinning.  In Figure \ref{coefs}, we display the coefficients that are greater than 0.05 across all five outcomes along with their 95$\%$ credible intervals. Different number of coefficients appear in each column of the plot that corresponds to each outcome, which is a result of the different levels of sparsity induced from the model. The resulting coefficients confirm existing knowledge in the literature of important covariates that predict these complications for geriatric patients. In addition, it suggests important procedures and medications that should be furthered flagged for patients to prevent higher risk of complications. Specifically, procedures for organ transplants, removal or insertion of a cardiac pacemaker, and heart valve procedures increase the risk of cardiac complications. Some procedures are inherently less risky across the surgical outcomes, including procedures on muscles and tendons, joint replacements that are not hip or knee, and procedures on the nose, mouth, and ears.  The number of medications patients take is strongly predictive of cardiac, pulmonary, and vascular complications, and whether they are on anticoagulants increases the risk of vascular and cardiac complications.  Risk factors for neurological complications, which includes dementia, are alcoholism, need for fluids and electrolytes, which indicates a nutritional deficiency, diabetes with complications, paralysis, and previous neurological problems. These align well with the literature on risk factors of dementia, providing further evidence that our model detects predictive covariates that are specific to the geriatric population.  In addition, an interesting feature of the chosen coefficients are their high correlation with one another. Typically in lasso, highly correlated coefficients are shrunk so that only one remains in the model. A nice feature in our model is that we can characterize patients more accurately regardless of how correlated the covariate space is, and provides a more accurate summary of important features. More importantly, these coefficients point to additional characteristics to better identify patients in the clinical setting.

%\end{minipage}

\section{Discussion}
We introduced the hierarchical infinite factor model that utilizes a hierarchical Dirichlet process weighting scheme as a sparsity-inducing transfer learning model. We contributed an easy-to-implement inference method and showed promising results that our method is effective at predicting surgical complications between unbalanced and sparse populations. Through simulation, we show that compared to state-of-the art baseline models, our model has better predictive accuracy and more accurate estimates of the coefficients, regardless of data size and type.  In addition, simulations show that HIFM flexibly models each population with its own factor loadings matrix that controls the number of factors needed to best explain the data. The resulting factor scores are a new representation of the data that diminishes the distributional differences between the populations, resulting in similar predictive performance regardless if one population is smaller than the other. 

Others in the literature have utilized transfer learning to improve prediction in health care settings. \citet{gong2015instance} proposed an instance weighting algorithm used in risk stratification models of cardiac surgery using a weighting scheme based on distances of each observation to the mean of the target distribution's predictors. \citet{wiens2014study} discussed the problem of using data from multiple hospitals to predict hospital-associated infection with \emph{Clostridium difficile} for a target hospital.   \citet{lee2012adapting} describe a method for transfer learning for the American College of Surgeon's National Surgical Quality Improvement Program (NSQIP) dataset, predicting mortality in patients after 30 days. Their methodology uses cost-sensitive support vector machines, first training the model on source data and next fitting the same model for the target data, but regularizing the model parameters toward that of the source model. While these approaches succeed in accomplishing positive transfer in their individual applications, their methods fail to learn the dependence structure underlying the observed data and do not provide any uncertainty quantification to the predicted outcomes. Our approach not only achieves positive transfer learning such that prediction is improved in the target task, but it also provides interpretable insights into potential phenotypes of patients that best explain those at risk for complications post-surgery. We show above that using this predictive tool compared to the current POSH heuristic increases the sensitivity of  death from 0.35 to 0.91. Improving sensitivity by almost a factor of 3 would have a huge impact for the geriatric patients at Duke. Implementing our proposed model in practice has the potential to save lives by either appropriately intervening on the patient or having further follow-up to decide whether the surgery is the right option for that patient.

While this work has focused on transfer learning between multiple populations, the model also shows promise as a sparsity inducing prior for single populations.  In future work, we aim to develop this model further in two directions. First, as an improved transfer learning model that better shares information across multiple populations. With such large imbalances between geriatric and the full population and with low signal in many of the variables, the model often struggles to model the local population accurately, leading to more noise and less accurate predictions. In addition, the data contain many binary variables that require transformations to use with our model. Another avenue for future work is to better address this binary data type to reduce the additional uncertainty added to the inference through the mapping of the binary variables into the continuous space. The second direction will be to explore this model further as a sparse factor model, without explicitly aiming to perform transfer learning. The properties proved in Section \ref{properties} hold for a single population, therefore providing potential for further development as a shrinkage prior. Lastly, we look further to testing and evaluating this model on additional applications in the health realm. If the HIFM is applied to new types of data, new properties in the feature space, such as group-specific covariates or different data structures will be of interest. 

Additionally, one could consider the Laplace distribution, or commonly known as the double-exponential distribution, as a prior for the factors, $\bm{f_i}$. Laplace distributed factors provide two additional features to the model: First, it induces sparsity through the factor distribution, which may improve model fit in sparse settings. Second, it provides an improvement to the indeterminacy problems that occur naturally with Gaussian factor models. We studied our model with Laplace distributed factors and found that it provided no additional benefit in the prediction for our particular application, but in other settings where identifiability is more of a concern, this is a reasonable alternative to the proposed model above.

Our work is a part of the continued effort to create a clinical platform to deliver individualized risk scores of complications at our university's health system for the purpose of triaging patients into preoperative clinics based on their underlying surgical risk. We plan to implement this framework directly into their electronic health system, so that clinicians will be able to assess the predicted complications directly through the patient's chart and treat the patient with suggested interventions that address the patient's increased risk. 
%Given that there is a limited amount of literature on causal inference for traffic safety evaluation, we focus on DID within the potential outcome framework for estimating the causal effect with the before-after safety data, and discuss DID methods based on outcome regression, propensity score weighting and their DR extension. The simulation studies illustrate that the proposed DR estimator has small bias as long as either the propensity score model or the crash frequency model are correctly specified, and therefore tends to provide more credible results.

%

\appendix
\section{Proofs of HIFM Properties} \label{App}

The following properties will be proven for each population, $l$, but we will drop the subscript for notational clarity.
\begin{align*}
\text{Proposition 1:} \operatorname{ If}\, (\Lambda, \Sigma) \sim \Pi_\Lambda \otimes \Pi_\Sigma,\, \operatorname{then}\\
\Pi_\Lambda \otimes \Pi_\Sigma(\Theta_\Lambda \times \Theta_\Sigma) = 1.
\label{prop1}
\end{align*}

\textit{Proof of Proposition \ref{prop1}: }
Because $\Pi_\Sigma(\Theta_\Sigma) = 1$ by definition of probability distributions, we only have to prove that $\Pi_\Lambda(\Theta_\Lambda) = 1$. 
We marginalize the distribution for lambda, $\lambda_{jh}|w_h, \phi_h \sim \text{Norm}(0, w_h \phi_{jh}^{-1})$, yielding a $t$ distribution with $v$ degrees of freedom with location and scale, $\lambda_{jh}|w_h \sim t_v(0, w_h)$. By the Cauchy-Schwartz inequality, 
\[ (\sum_{h=1}^{\infty} \lambda_{rh}\lambda_{sh})^2 \le (\sum_{h=1}^{\infty} \lambda_{rh}^2)(\sum_{h=1}^{\infty}\lambda_{sh}^2) \le \max_{1 \le j \le p} (\sum_{h=1}^{\infty}\lambda_{jh}^2)^2\]
Therefore, 
\[(\sum_{h=1}^{\infty} \lambda_{rh}\lambda_{sh}) \le  \max_{1 \le j \le p} (\sum_{h=1}^{\infty}\lambda_{jh}^2) \]
Let ${M}_j = (\sum_{h=1}^{\infty}\lambda_{jh}^2)$ and ${M} =  \max_{1 \le j \le p} {M}_j$, where all elements of $\Lambda \Lambda^T$ are bounded in absolute value by $M$.
\begin{align*}
E({M}_j) &= \sum_{h=1}^\infty E[E(\lambda_{jh}^2|w_h)] \\
&= \sum_{h=1}^\infty E[w_h \frac{v}{v-2}] \\
&= \sum_{h=1}^\infty \frac{v}{v-2} E[E(w_h|\pi_{0h})] \\
&= \frac{v}{v-2} \alpha_l \sum_{h=1}^\infty  E[\pi_{0h}] \\
&= \frac{v}{v-2} \alpha_l  \lim_{K \rightarrow \infty}  E[\sum_{h=1}^K \pi_{0h}]\\
&= \frac{v}{v-2} \alpha_l < \infty. 
\end{align*}

Therefore, $E(M) \le  \sum_{j=1}^P E(\text{M}_j) < \infty$.  Therefore, $M$ is finite almost surely. It follows that $\Pi_\Lambda \otimes \Pi_\Sigma(\Theta_\Lambda \times \Theta_\Sigma) = 1$.   //\\

\textit{Proof of Theorem \ref{th2}: }
This theorem is proved by first showing the following properties as defined in Proposition \ref{prop2} and Lemma \ref{lem3} from \cite{bhattacharya2011sparse}:

\begin{prop}\label{prop2}
If $\Omega_0$ is any $p \times p$ covariance matrix and $B^\infty_\epsilon(\Omega_0)$ is an $\epsilon$-neighbor of $\Omega_0$ under sup-norm, then $\Pi\{B^\infty_\epsilon(\Omega_0)\}>0$ for any $\epsilon>0$. 
\end{prop}

The proof of Proposition 2 follows closely to the proof in \cite{bhattacharya2011sparse}. Let $\Lambda_*$ be a $p\times k$ matrix and $\Sigma_0 \in \Theta_\Sigma$ such that $\Omega_0 = \Lambda_* \Lambda_*^T + \Sigma_0$. Set $\Lambda_0 = (\Lambda_* : 0_{p \times \infty})$, then $(\Lambda_0, \Sigma_0) \in \Theta_\Lambda \times \Theta_\Sigma$, with $g(\Lambda_0, \Sigma_0)=\Omega_0$. Fix $\epsilon>0$, and choose $\epsilon_1>0$ such that $(2M_0+1)\epsilon_1+\epsilon_1^2 <\epsilon$, where $M_0 = \max_{1\le j \le p} (\sum_{h=1}^{\infty}\lambda_{jh}^{0 2})^{1/2}$. By Lemma 2 in \cite{bhattacharya2011sparse}, $g\{B_{\epsilon_1}(\Lambda_0, \Sigma_0)\} \in B^\infty_\epsilon(\Omega_0)$, and therefore $B_{\epsilon_1}(\Lambda_0, \Sigma_0) \in g^{-1}\{B^\infty_\epsilon(\Omega_0)\} \ge \Pi_\Lambda \otimes \Pi_\Sigma\{B_{\epsilon_1}(\Lambda_0, \Sigma_0)\}$. It is obvious that $\Pi_{\Sigma}\{\Sigma: d_\infty(\Sigma, \Sigma_0)<\epsilon_1\}>0$, which leaves us only to show that $\Pi_\Lambda = \{\Lambda: d_2(\Lambda, \Lambda_0)<\epsilon_1\}>0$. The next steps is where we adapt the remainder of the proof to our prior. 

\begin{align}
\text{pr}\{d_2(\Lambda, \Lambda_0)<\epsilon_1\} &= \text{pr}\{\sum_{j=1}^p \sum_{h=1}^\infty (\lambda_{jh} - \lambda_{jh}^0)^2 < \epsilon_1^2 \}\\
&\ge \text{pr}\{\sum_{h=1}^\infty (\lambda_{jh} - \lambda_{jh}^0)^2 < \epsilon_1^2/p, \, j=1,...,p \}\\
&= E_w[\prod_{j=1}^p \text{pr}\{\sum_{h=1}^\infty (\lambda_{jh} - \lambda_{jh}^0)^2 < \epsilon_1^2/p | w_{lk}\} ] >0.
\end{align}

This is shown from Lemma \ref{lem3}.

\begin{lemma}\label{lem3}
Fix $1\le j \le p$. For any $\epsilon>0$, $pr\{\sum_{h=1}^\infty (\lambda_{jh} - \lambda_{jh}^0)^2 <\epsilon/2. \sum_{h=H+1}^{\infty} \lambda_{jh}^2 <\epsilon/2|w_{h} \} >0$ almost surely. 
\end{lemma}

\textit{Proof of Lemma \ref{lem3}: }
\begin{align*}
\text{pr}\{\sum_{h=1}^\infty (\lambda_{jh} - \lambda_{jh}^0)^2 < \epsilon | w_{h}\}\ge \\
\text{pr}\{\sum_{h=1}^H (\lambda_{jh} - \lambda_{jh}^0)^2 < \epsilon/2,  \sum_{h=H+1}^{\infty} \lambda_{jh}^2 <\epsilon/2|w_{h}\} \\
=\text{pr}\{\sum_{h=1}^H (\lambda_{jh} - \lambda_{jh}^0)^2 < \epsilon/2 | w_h\} \, \text{pr}\{\sum_{h=H+1}^{\infty} \lambda_{jh}^2 <\epsilon/2|w_{h}\}
\end{align*}
The latter probability goes to 1 as $H \rightarrow \infty$, as shown from Theorem 1. Therefore, we can find an $H_0>k$ such that $\text{pr}(\sum_{h=H_0+1}^{\infty} \lambda_{jh}^2 <\epsilon/2)>0$. This implies that $\text{pr}(\sum_{h=H+1}^{\infty} \lambda_{jh}^2 <\epsilon/2|w_{h}) > 0$ almost surely. Therefore, $\text{pr}\{\sum_{h=1}^H (\lambda_{jh} - \lambda_{jh}^0)^2 < \epsilon/2|w_h\}>0$ almost surely for any $H<\infty$. 

By proving the above Lemmas and Theorems for our prior, the proof of Theorem \ref{th2} follows exactly from \citet{bhattacharya2011sparse}. //

\section{Inference - Full Sampler}\label{inf}
The following steps provides the full sampling scheme for the HIFM. For each patient $i$, we draw from the following multivariate normal distribution.  
\[(f_i|\Lambda_l, \Sigma_l, x_i) \sim \text{N}_{k}(\mu_{f_l}, \Sigma_{f_l}) \]
where $\mu_{f_l} = x_i \Sigma_l^{-1}\Lambda_l (I_k+\Lambda_l'\Sigma_l^{-1}\Lambda_l)^{-1}$ and $\Sigma_{f_l} =(I_k+\Lambda_l'\Sigma^{-1}\Lambda_l)^{-1}$ are the posterior mean and covariance, respectively. $\Lambda_l$ represents the loadings matrix for the $l$th population

Next, sample the $j$th row of $\Lambda_l$, where $\lambda_{lj}$ denotes the $l$th population loading matrix at row $j$. We denote the data in group $l$ by subscripting the local parameters with an $l$, where $x_{lj}$ represents the $j$th column of data in group $l$, and $F_l$ is the factor matrix with rows contained in group $l$. 
\[(\lambda_{lj}|x_j, F_l, w_l) \sim N_{k}(\mu_{\lambda_{lj}}, \Sigma_{\lambda_{lj}}) \,,\]
where the posterior mean $\mu_{\lambda_{lj}} = F_l'X_{lj}\sigma_{lj}^{-2}(D_{lj}^{-1} + \sigma^{-2}_{lj} F_l'F_l)^{-1}$ and the covariance $\Sigma_{\lambda_{lj}} = (D_{lj}^{-1} + \sigma^{-2}_{lj} F_l'F_l)^{-1}$, with $D_{lj}^{-1} = \text{diag}(\phi_{j1}/w_{l1},..., \phi_{jk}/w_{lk})$.

Next, we draw $\sigma^2_{lj}$ or the variance corresponding to the $j$th covariate for population $l$.
\[(1/\sigma^2_{lj}|\lambda_l F_l, x_{lj}) \sim {\rm Gamma}(a_{\sigma^2_{lj}}, b_{\sigma^2_{lj}}) \,, \]
where $a_{\sigma^2_{lj}} = v/2 + n_l/2$ and $b_{\sigma^2_{lj}} = 1/2(v + (x_{lj} - F\lambda_{lj})^T(x_{lj} - F_l\lambda_{lj})$.

For element in row $j$, column $h$ of precision parameters, $\phi$, we update using a Gamma distribution.
\[\phi_{jh} \sim {\rm Gamma}(a_{\phi_{jh}},b_{\phi_{jh}} ) \,, \]
where $a_{\phi_{jh}} = \tau/2+L/2$ and $b_{\phi_{jh}} = \tau/2+\sum_{l=1}^L \frac{\lambda_{ljh}}{2w_{lh}}$.

The weight parameters $\bm{w_l}$  are updated with a closed form draw from the generalized inverse-gaussian distribution for each $h$th element of $w_l$:
\[(w_{lh}|\lambda_l, \pi_0, \alpha_l) \sim {\rm GIG}(p=p_{w_{lh}}, a=a_{w_{lh}}, b= b_{w_{lh}})) \,, \]
where $p_{w_{lh}} = \alpha_l \pi^0_h - p/2$, $a_{w_{lh}}=2$, and $b_{w_{lh}}=(\lambda_{lh}'\Phi_h\lambda_{lh})$. $\Phi_h = {\rm diag}(\phi_{h1},..,\phi{hp})$.

To update $\pi_0$, we first propose $\theta_{0h}^* \sim \text{Gamma}(\theta_{0h}^{t-1} \cdot C, C)$, which gives a mean of $\theta_{0h}^{t-1}$ and a variance of $\theta_{0h}^{t-1}/ C$ which allows tuning using the constant, C. We then normalize the $\bm{\theta_0^*}$, such that $\pi_0^* = \frac{\bm{\theta_0^*}}{\sum_{h=1}^k \bm{\theta_{0h}^*}}$,
and accept  $\pi_0^*$ based on the acceptance ratio: 
\[ A(\pi_0^*|\pi_0^{t-1})=\min \left(1,{\frac {P(\pi_0^*|w_1,..., w_l)}{P(\pi_0^{t-1}|w_1,..., w_l)}}{\frac {g(\pi_0^{t-1}|\pi_0^*)}{g(\pi_0^*|\pi_0^{t-1})}}\right) \]

\section*{Acknowledgements}
The authors are grateful to Fan Li for providing helpful suggestions on the overall discussion of this work. We also acknowledge the Duke Institute for Health Innovations, specifically Kristin Corey, Sehj Kashyap, and Mark Sendak, for providing the data and clinical insights into the findings of our work.

\bibliographystyle{jasa3}
\bibliography{AOAS_HIFM_2018}

\end{document}